# Changing EDSS progression in placebo cohorts in relapsing MS: A systematic review and meta-regression

**Christian Röver[1], Richard Nicholas[2], Sebastian Straube[3], Tim Friede[1]**

[1]Department of Medical Statistics, University Medical Center Göttingen, Göttingen, Germany

[2]Imperial College Healthcare NHS Trust, London, UK

[3]Division of Preventive Medicine, University of Alberta, 5-30F University Terrace, 8303-112 Street, Edmonton, AB, T6G 2T4, Canada

*Corresponding author:*

Prof Tim Friede, University Medical Center Göttingen, Department of Medical Statistics, Humboldtallee 32, D-37073 Göttingen; Tel +49-551-39 4991; Fax +49-551-39 4995; tim.friede@med.uni-goettingen.de

Figures: 2 Tables: 2

Number of references: 19

Character count for title: 88

Word count: 3048 (including 236 for the abstract)

## Abstract

**Background:** Recent systematic reviews of randomised controlled trials (RCTs) in relapsing multiple sclerosis (RMS) revealed a decrease in placebo annualized relapse rates (ARR) over the past two decades. Furthermore, regression to the mean effects were observed in ARR and MRI lesion counts. It is unclear whether disease progression measured by the expanded disability status scale (EDSS) exhibits similar features.

**Methods:** A systematic review of RCTs in RMS was conducted extracting data on EDSS and baseline characteristics. The logarithmic odds of disease progression were modelled to investigate time trends. Random-effects models were used to account for between-study variability; all investigated models included trial duration as a predictor to correct for unequal study durations. Meta-regressions were conducted to assess the prognostic value of a number of baseline variables.

**Results:** The systematic literature search identified 39 studies, including a total of 19,714 patients. The proportion of patients in placebo controls experiencing a disease progression decreased over the years ($p<0.001$). Meta regression identified associated covariates including the size of the study and its duration that in part explained the time trend. Progression probabilities tended to be lower in the second year compared to the first year with a reduction of 24% in progression probability from year 1 to year 2 ($p=0.014$).

**Conclusion:** EDSS disease progression exhibits similar behaviour over time as the ARR and point to changes in trial characteristics over the years, questioning comparisons between historical and recent trials.

## Introduction

Recent systematic reviews of placebo groups in randomised controlled trials (RCTs) in relapsing multiple sclerosis (RMS) suggest a decrease in annualized relapse rates (ARRs) over calendar time [1-3] as well as a decrease in relapse rates over the course of the study [4]. These changes appear to relate to the changing eligibility criteria and populations entering trials and regression to the mean effects [3]. Pre-trial ARR and mean baseline age were independently identified as predictors for on-trial ARR in a smaller number of phase III trials with at least 18 months follow-up by Stellmann et al. [5]. Recently regression to the mean in MRI lesion counts was identified and quantified in a systematic review and meta-analysis [6].

Disability outcomes in multiple sclerosis (MS) are a key component that regulators have identified as the principal target for an increasing range of therapies targeting the underlying disease process [7]. The commonly used method of disability measurement in trials is the extended disability status scale (EDSS) [8].

In this paper we aim to investigate whether a decrease in placebo ARRs observed in randomized controlled trials in RMS is also present in EDSS progression and, if yes, whether it can also be explained by changes in patient populations and design characteristics. Furthermore, we will assess placebo controls of RCTs for regression to the mean effects in EDSS progression.

## Methods

### Systematic literature search

A recently conducted systematic literature search of placebo controlled randomised trials in RMS [3] was updated by searching PubMed with the aim of identifying placebo-controlled, double-blind RCTs in MS where all or some of the patients had a relapsing form of the disease and that reported data on pre-trial and on-trial ARR as well as on pre-trial and on-trial EDSS. To update the previous systematic review we searched for articles published from 2011 onwards with the search terms “multiple sclerosis”, “relapse rate” and “placebo”. All abstracts were independently screened by two reviewers. The search was performed February 20$^{th}$, 2015. If one reviewer suggested the full paper be examined after reading the abstract, the full paper was considered. For a trial to be included in this systematic review it had to be randomised, single or double blind, and placebo-controlled, with at least some of the trial participants having RMS. Trials had to assess the efficacy of disease modifying drugs (i.e. not assessing symptomatic therapies), and report data on both clinical relapses and EDSS. We excluded cross-over trials and studies where patients in the control group received a form of active treatment (add-on therapy).

### Data extraction

The following data were extracted by one reviewer and verified by another:

- publication date, treatment groups and corresponding numbers of patients, and duration of follow-up,
- the proportion of patients exhibiting a worsening in EDSS,
- the ordinates at years 1 and 2 of Kaplan-Meier curves of confirmed EDSS progression, or equivalent tabulated data.

### Data analysis

For the purpose of all analyses of temporal trends, the year and month of publication and the study durations were used. EDSS progression was analyzed based on the *logarithmic odds* (*log-odds*) of disease progression; these result as $log\left(\frac{\hat{p}}{1-\hat{p}}\right)$, where $\hat{p}$ is the fraction of progressing patients. Corresponding standard errors are calculated based on a binomial model as $\hat{\sigma} = \sqrt{\frac{1}{k} + \frac{1}{N-k}}$ , where $N$ is the total number of patients, out of which $k$ have progressed. In the meta-regression analyses we utilized linear regression methods, accounting for the individual standard errors. Random-effects models were used in order to account for potential between-study variability, and all investigated models included trial duration as a predictor in order to correct for unequal study durations [9]. Between-study heterogeneity $\tau^2$ was estimated using the Mandel-Paule method [10] which is reported with the p-values of the chi-square test of heterogeneity. For the multivariate regression, we used the *Bayesian information criterion (BIC)* to determine the best-fitting model among all possible subsets of predictors [11]. For the predicted means 95% confidence intervals were calculated. We used random-effects meta-analysis to investigate the probability of EDSS progression during first and second year of a study. In estimating the standard errors of the log risk ratios we neglected correlations between the risk estimates (the empirical fractions of patients are positively correlated, and so we err on the conservative side here, underestimating the uncertainty in differences or ratios).

## Results

### Studies identified

The systematic literature search identified 39 studies, including a total of 19,714 patients, of which 6,947 received placebo. The cumulative observation time amounts to 31,368 patient-years, with a contribution of 11,163 patient-years from placebo treated patients. The study selection process is illustrated in the flow chart in Fig. 1. The study characteristics are summarized in Table 1. Disease progression can be defined in a number of ways, of the 39 studies, disability worsening of one point or more on the EDSS scale was used in 34 (87%) and progression confirmation of 3 months or more was used in 23 (59%).

**Decreasing placebo EDSS progression rates over the past two decades**

Figure 1 shows the fractions of progressing patients as reported in different studies' placebo groups over the years. The chances of progression of course also depend on the trial duration ("shorter" and "longer" studies are also indicated by different symbols), but even after accounting for the individual follow-up times, the regression analysis indicates a statistically significant effect of the publication year with the odds for disease progression decreasing by 31% (95% CI [17%, 42%]) within a decade (p<0.001; between-trial heterogeneity $\tau^2$= 0.15, p<0.001).

**Predictors of EDSS disease progression**

Table 2 shows the results from univariate regression analyses using the remaining covariates (excluding the publication year). For each of the predictors investigated, the odds ratio, which is the multiplicative change in odds for EDSS progression for every unit increase in the predictor, is shown along with its 95% confidence interval and p-value. Note that all investigated models included follow-up time as a covariate. For each analysis, the value of $\tau^2$, the random effect accounting for between-study heterogeneity is shown as well, along with the relative reduction in $\tau^2$ compared to the model that only uses follow-up time as a predictor. With larger numbers of placebo patients the progression probability decreases (p=0.0004). As would have been expected, the progression probability increases with the length of the study (p=0.023 for <=1 vs. >1 year; p<0.0001 for linear trend).

The model selection result is also shown in Table 2. The variables included in the final model relate to study size (number of placebo patients: $p<0.001$), study duration (follow-up duration in years: p=0.26; study longer 1 year: p=0.018) and baseline disease status (baseline ARR p=0.69; mean baseline EDSS: p=0.18) and confirmation of disease progression (p=0.25).

### Decreasing EDSS disease progression over follow-up time

Figure 2 illustrates the probabilities of progression during the first and second years of follow-up for the 13 studies where relevant data could be extracted. Overall progression probabilities tend to be lower in the second year, with the exception of two small studies from the 1990's. The combined risk ratio comparing the progression probability from year 2 to year 1 is 0.76 (95% CI [0.62; 0.94], p=0.014; between-trial heterogeneity $\tau^2$=0.057, p=0.0085) which translates to a reduction of 24% in progression probability from year 1 to year 2. Looking at combined progression probabilities from random-effects meta-analyses, chances are 17.7% during the first year, and 13.0% during the second year. These studies used confirmed disease progression as endpoint. When considering only the eight most recent studies published during the last decade, the numbers change slightly to 17.1% and 11.5% during first and second year, respectively.

## Discussion

For the key clinical outcome measure of disease progression in MS [7], this study has confirmed in relapsing MS trial placebo groups the rates of disease progression reduce with the more recent publication year of the study and between the first and second year of the study.

Increasing study size and decreasing length of the study were significant in reducing rates of disease progression seen in studies with later publication year. Study size has already been shown to be highly correlated with publication year for ARR [3]. However a reduced ARR with later publication date was associated with a shorter study length whereas an increased rate of EDSS progression rate was associated with increased study length. For ARR this is because the earlier studies were longer and the study populations of the earlier trials had higher disease activity whereas for disease progression the increase with length of study is due to the way in which cumulative progression is calculated i.e those who progress in the first year are added to those who progress subsequently. Thus both these features this may well be simply confounding factors and unlike in the case of ARR we did not find any other subject or study design features explaining this trend [3].

We have also found that the rates of disease progression reduce from the first to the second year implying as with ARR rates [4] there is a regression to mean effect. However, as with the time-to-first relapse endpoint an alternative explanation for the apparent time-dependence in the time-to-disease progression data could be between-patient heterogeneity [12].

Although correlations on a patient level between MRI outcomes, relapse and EDSS progression are small [13,14], known as the clinico-radiological paradox [15], the trends over time on a population level are quite similar in these three measures. Given the similar relationship what we may be seeing is the impact of the early phase of MS relapses and MRI activity on the disability score. The EDSS has acknowledged limitations [16-18] and levels of disability especially in early MS are very variable [19] as relapses are known to bias the assessment of disability. Here disability progression was defined predominantly using a one point increase in the EDSS combined with 3 months confirmation. Time to 3 months confirmed disability progression is known to be more susceptible to the impact of relapses than measures that use a longer period to confirm a change in disability [7,18]. In summary, EDSS disease progression as used here exhibits similar behaviour over time as the ARR and point to changes in trial characteristics over the years, questioning comparisons between historical and recent trials.

## Acknowledgements

RN is grateful for support from the NIHR Biomedical Research Centre.

**Table 1:** Baseline characteristics of all randomised patients and the patients in the placebo groups in the 39 randomised, controlled trials included in this systematic review. *N* denotes the number of treatment arms for which the corresponding figures could be extracted.

| | **Placebo groups** | | **All treatment groups** | |
|---|---|---|---|---|
| | ***N*** | ***Median (range)*** | ***N*** | ***Median (range)*** |
| **number of patients** | 39 | 99 (9 – 556) | 97 | 123 (9 – 943) |
| **study duration (years)** | 39 | 2.00 (0.46 – 5.00) | 97 | 1.85 (0.46 – 5.00) |
| **mean pre-trial EDSS** | 36 | 2.68 (1.94 – 5.05) | 91 | 2.69 (1.86 – 6.05) |
| **mean on-trial EDSS** | 23 | 2.88 (1.88 – 5.59) | 55 | 2.79 (1.87 – 6.50) |
| **EDSS progressing proportion** | 39 | 0.23 (0.04 – 0.50) | 97 | 0.18 (0.04 – 0.76) |
| **mean pre-trial ARR** | 32 | 1.40 (0.87 – 2.10) | 80 | 1.37 (0.75 – 2.10) |
| **mean on-trial ARR** | 39 | 0.81 (0.22 – 1.80) | 97 | 0.54 (0.14 – 1.80) |

**Table 2:** Results of univariate and multivariate regression aiming at explaining the probability of EDSS progression. Regression coefficients relate to the logarithmic odds of progression. $\tau^2$ denotes the unexplained between-study heterogeneity, and the reduction percentages relate to the model including only follow-up duration as a predictor (which is also included in all univariate models). The multivariate model was selected based on the *Bayesian information criterion* (BIC).

| | **univariate** | | | | **multivariate** | | | |
|---|---|---|---|---|---|---|---|---|
| **variable** | ***odds ratio*** | ***95% CI*** | ***p-value*** | ***$\tau^2$ (red. %)*** | ***odds ratio*** | ***95% CI*** | ***p-value*** | ***$\tau^2$ (red. %)*** |
| **number of placebo patients** | 0.99839 | 0.99750, 0.99928 | 0.00040 | 0.405 (32.2) | 0.99744 | 0.99607, 0.99880 | 0.00024 | 0.366 (44.5) |
| **long study (>1 year)** | 1.852 | 1.087, 3.155 | 0.023 | 0.457 (13.7) | 1.877 | 1.156, 3.157 | 0.018 | 0.366 (44.5) |
| **Oxford quality score** | 0.777 | 0.602, 1.003 | 0.053 | 0.468 (9.4) | | | | |
| ***confirmed* progression (yes)** | 0.709 | 0.475, 1.058 | 0.092 | 0.475 (6.4) | 1.369 | 0.799, 2.347 | 0.25 | 0.366 (44.5) |
| **mean baseline age (years)** | 0.9783 | 0.9192, 1.0412 | 0.49 | 0.482 (3.9) | | | | |
| **eligibility criteria: number of words** | 0.999042 | 0.997634, 1.000452 | 0.18 | 0.482 (2.8) | | | | |
| **mean baseline EDSS** | 1.0383 | 0.7749, 1.3913 | 0.80 | 0.485 (2.0) | 0.822 | 0.619, 1.092 | 0.18 | 0.366 (44.5) |
| **mean baseline MS duration (years)** | 0.9791 | 0.8912, 1.0756 | 0.66 | 0.490 (0.6) | | | | |
| **followup duration (years)** | 1.614 | 1.312, 1.984 | 0.0000058 | 0.492 (0.0) | 1.171 | 0.889, 1.543 | 0.26 | 0.366 (44.5) |
| **eligibility criteria: number** | 0.99475 | 0.98193, 1.00774 | 0.43 | 0.495 (-1.4) | | | | |
| **number of treatment arms** | 0.788 | 0.577, 1.078 | 0.14 | 0.498 (-2.4) | | | | |
| **baseline ARR** | 1.290 | 0.663, 2.511 | 0.45 | 0.528 (-15.2) | 0.883 | 0.477, 1.634 | 0.69 | 0.326 (56.1) |

**Figure 1:** The fractions of patients with progressing EDSS status over the years. The chances of progression also depend on the study duration (you can see that shorter studies have smaller fractions) but even after accounting for the duration, the decreasing trend remains statistically significant ($p<0.0001$). The red line shows the estimated regression line for a trial duration of 1 year.

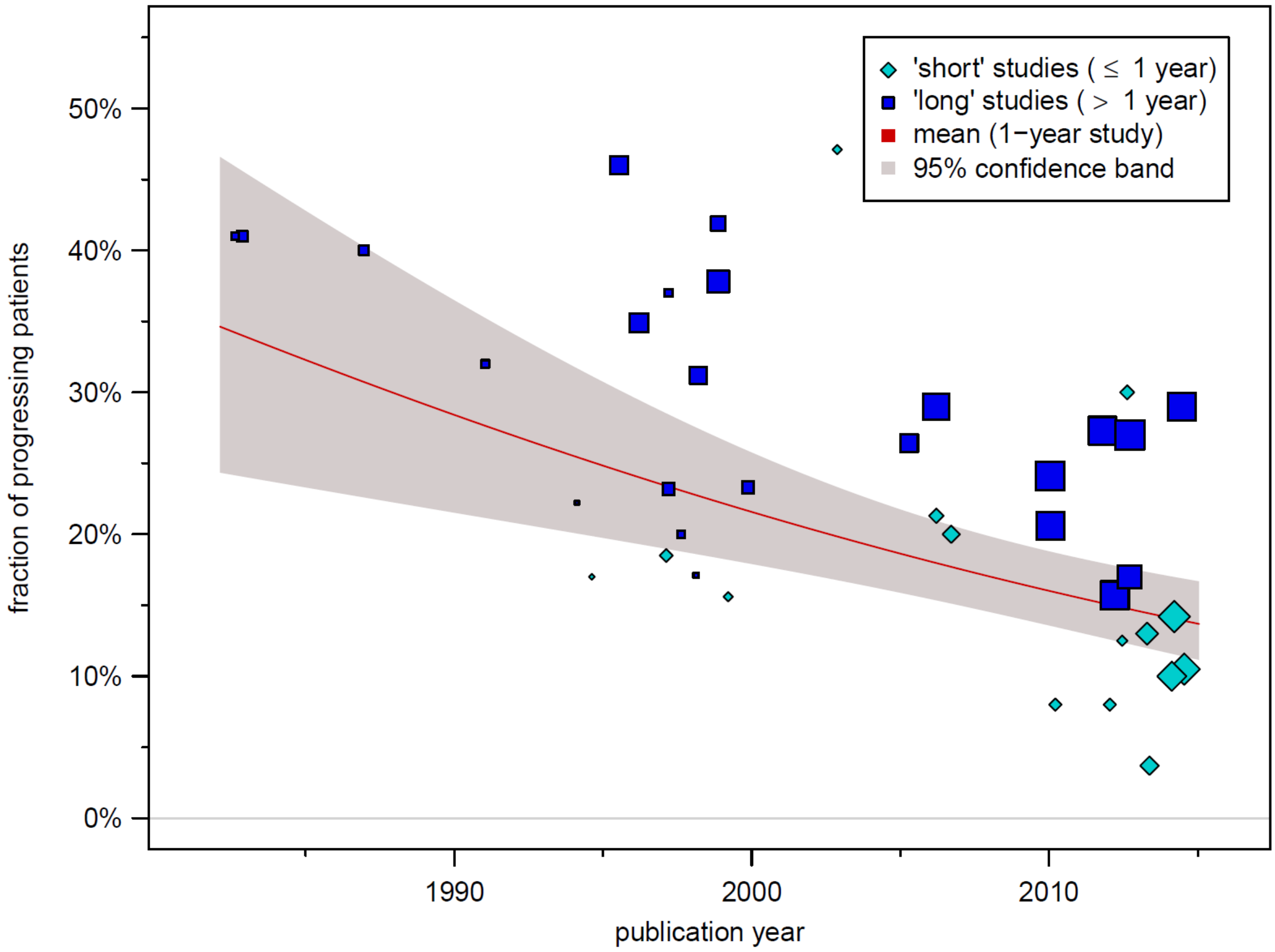

**Figure 2:** The fractions of patients with progressing EDSS in the first and second year of study, for the 13 studies of at least 2 years duration, and where the data was provided. Connecting lines indicate the rates for the two subsequent years, line widths are proportional to study sizes (numbers of patients *N*). The weighted average (weighted by study size *N*) decreases from 17.7% to 13.0% from first to second year. [17.1% to 11.5% for 8 most recent post-2000 studies].

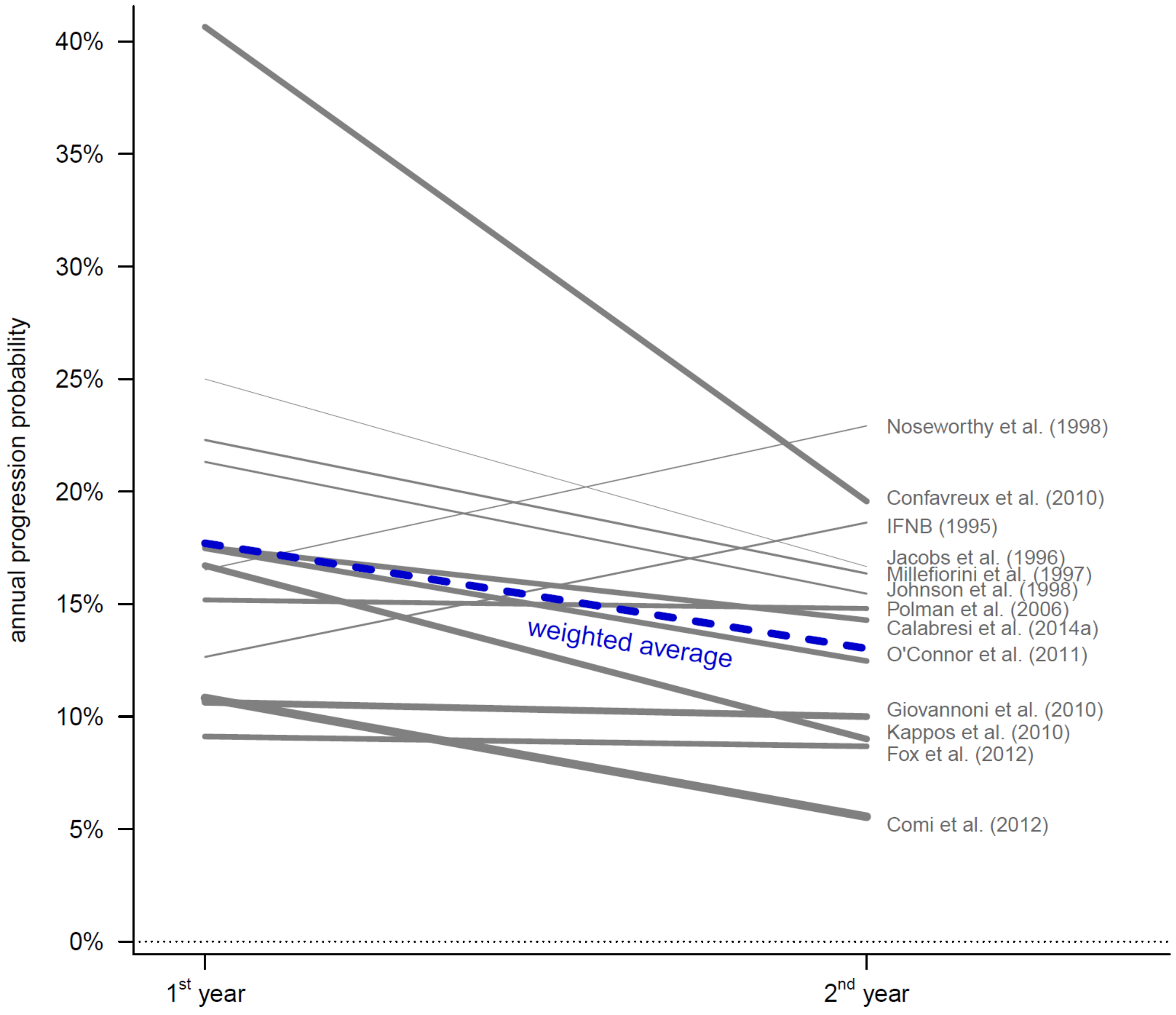

**Figure S1:** The PRISMA flow chart illustrating the systematic literature review.

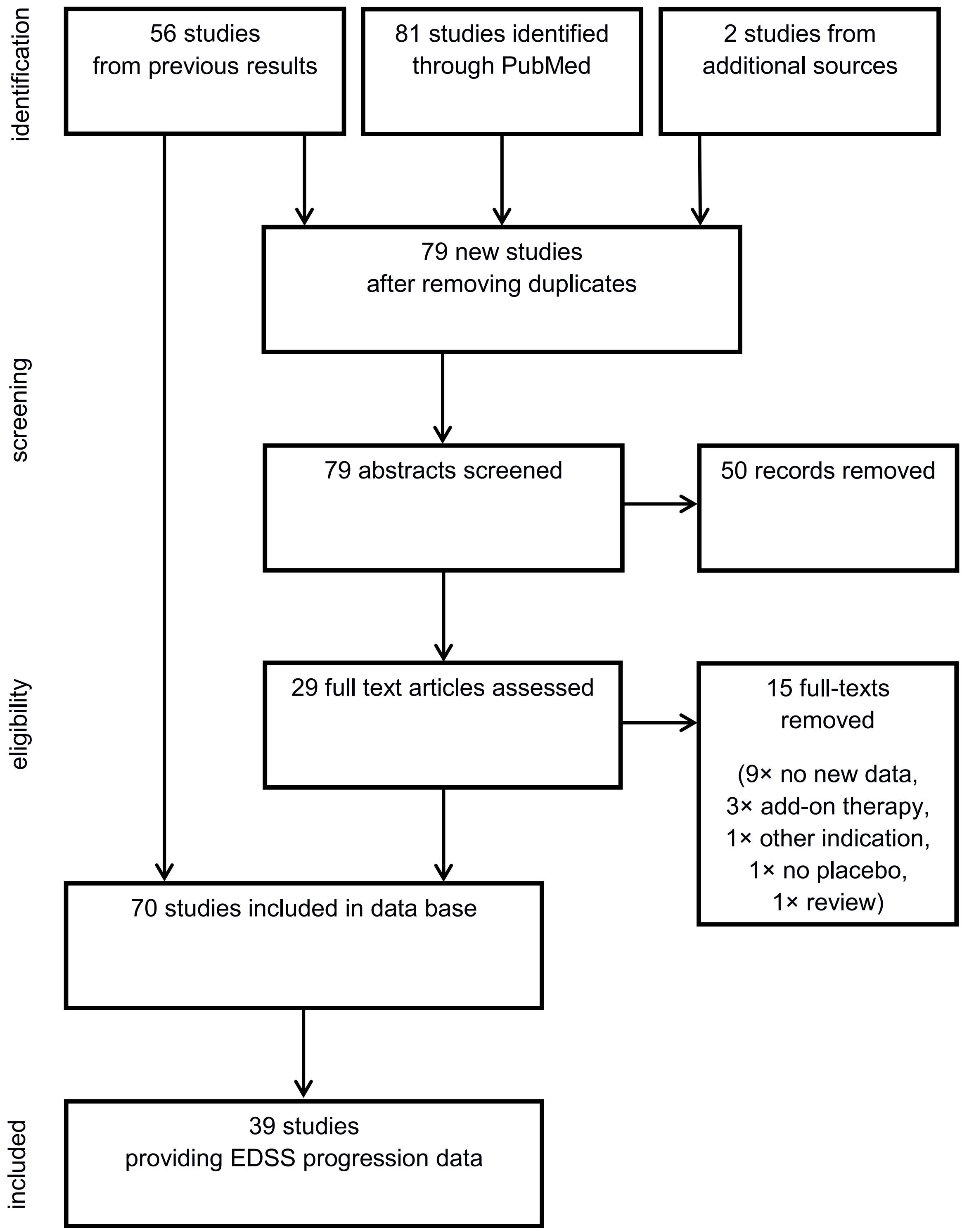